\documentstyle[prd,aps,floats,epsf,psfrag]{revtex}

\newcommand{\eq}[1]{equation~(\ref{#1})}
\newcommand{\eqs}[2]{equations~(\ref{#1}) and~(\ref{#2})}
\newcommand{\Eqs}[2]{Equations~(\ref{#1}) and~(\ref{#2})}
\newcommand{\Eq}[1]{Equation~(\ref{#1})}

\newcommand{\be}{\begin{equation}}
\newcommand{\ee}{\end{equation}}

\def\beq {\begin{equation}}
\def\eeq {\end{equation}}
\def\beqa {\begin{eqnarray}}
\def\eeqa {\end{eqnarray}}

\def\At {\dot{A}}
\def\Bt {\dot{B}}
\def\Pt {\dot{\phi}}

\def\Att {\ddot{A}}
\def\Btt {\ddot{B}}
\def\Ptt {\ddot{\phi}}

\def\Az {A'}
\def\Bz {B'}
\def\Pz {\phi'}

\def\Bzz {B''}
\def\Pzz {\phi''}

\def\Btz {\dot{B}'}
\def\ha {\frac{1}{2}}

\begin{document}

\draft

\renewcommand{\topfraction}{0.99}
\renewcommand{\bottomfraction}{0.99}
\twocolumn[\hsize\textwidth\columnwidth\hsize\csname 
@twocolumnfalse\endcsname

\title{ Preheating and the Einstein Field Equations} 
\author{Matthew Parry and Richard Easther}
\address{Department of Physics,  Brown University, 
Providence, RI 02912, USA.}
\maketitle
\begin{abstract}
We inaugurate a framework for studying preheating and parametric
resonance after inflation without resorting to any approximations,
either in gravitational perturbation theory or in the classical
evolution of the field(s). We do this by numerically solving the
Einstein field equations in the post-inflationary universe.  In this
paper we show how to compare our results to those of gauge invariant
perturbation theory. We then verify Finelli and Brandenberger's
analysis (hep-ph/9809490) of super-horizon modes in $m^2\phi^2$
inflation, showing that they are not amplified by resonant effects.
Lastly, we make a preliminary survey of the nonlinear couplings
between modes, which will be important in models where the primordial
metric perturbations undergo parametric amplification.
\end{abstract}
\pacs{PACS numbers: 98.80.Cq,\quad 04.25.Dm \qquad BROWN-HET-1143  }]

\vskip 5mm

\noindent {\bf I. Introduction}  
Postulating a period of nearly exponential growth in the primordial
universe, inflationary cosmology \cite{Guth1981a} solved many problems
which plagued previous models of the big bang.  Immediately after
inflation the universe is dominated by the inflaton, the scalar field
whose evolution controls the dynamics of the inflationary era.  The
inflaton energy density is then converted into thermalized, ``normal''
matter, a process called {\it reheating\/}.  The first predictions
\cite{AbbottET1982a,AlbrechtET1982b,DolgovET1982a} were that the
universe would be reheated to a temperature several orders of
magnitude (or more) below the inflationary scale.  Subsequently, it
has been realized that non-equilibrium, non-perturbative processes can
drive explosive particle production at the end of inflation 
\cite{TraschenET1990a,%
KofmanET1994a,%
ShtanovET1995a,%
Kaiser1996a,%
KhlebnikovET1996a,%
KolbET1996a,%
KhlebnikovET1997a,%
KhlebnikovET1997b,%
BoyET1996,%
ProkopecET1996a,%
BoyET1997,%
RamseyET1997,%
KofmanET1997a,%
GreeneET1997a,%
GreeneET1997b}.
This initial era, dubbed {\it preheating\/}, is governed by
Mathieu-like equations, and the rapid particle creation associated
with the instability bands of these equations is often termed
{\it parametric resonance\/}.

Developing a full understanding of preheating has been a long and arduous
task. There were two main complications. Firstly the difference between
parametric resonance in Minkowski space and in an expanding universe is
profound. In the latter case, the instability bands may be time-dependent
and so a particular mode may pass in and out of resonance.  Secondly, the
back-reaction of the created particles onto the other fields must become
significant; and indeed the parametric resonance will eventually
terminate. With a few exceptions though, (see e.g. 
\cite{KodamaET1996a,%
NambuET1996a,%
BassettET1998a,%
FinelliET1998a}),
spacetime was described by the usual Friedman Robertson Walker metric. 
Although the scale factor could be determined self-consistently (see e.g.
\cite{RamseyET1997}), the possibility of resonant growth of {\it metric\/}
perturbations was not considered.  However, in a recent paper Bassett {\it
et al.\/} \cite{BassettET1998a} highlight the importance of couplings
between perturbations in the fields and in the metric.  As they point out,
metric perturbations are sources for the field perturbations, and {\it
vice versa}, so it cannot be assumed that the role of metric perturbations
in preheating is only slight (see also
\cite{KodamaET1996a,NambuET1996a,FinelliET1998a}).  During preheating, in
this framework, parametric resonance will necessarily amplify metric
perturbations and so nonlinear gravitational effects will become
significant. 

In this paper, we abandon the perturbative approach entirely and study
the post-inflationary era by numerically solving the inhomogeneous,
nonlinear Einstein field equations, and derive the connection between
our results and the predictions of perturbation theory.  We then
confirm the conclusions of Finelli and Brandenberger
\cite{FinelliET1998a} who use gauge invariant perturbation theory to
show that super-horizon modes are not parametrically amplified after
$m^2 \phi^2$ inflation. Lastly, we use the $m^2\phi^2$ theory as a toy
model to explore the consequences of having significantly amplified
modes in the post-inflationary universe. In particular, we see how
power can be transferred from modes with large amplitudes to all other
modes.  The mode-mode coupling thus appears to broaden the impact of
resonance in the early universe.  This is the first time in the study
of preheating that the mode-mode coupling of the field {\it and}
metric perturbations has been calculated in a fully relativistic
manner.  In $m^2\phi^2$ inflation, metric perturbations are not
amplified so we inject a large perturbation via our initial
conditions. In future work we will give a nonlinear treatment of
inflationary models such as $\lambda\phi^4$ and $g^2\phi^2\chi^2$,
where the field perturbations are amplified by resonant effects, and
study the role of mode-mode coupling in these systems.

\vskip 5mm

\noindent {\bf II. Metric and Field Equations}  We consider scenarios
whose initial state corresponds to a spatially flat Friedman Robertson
Walker universe with a perturbation consisting of a single Fourier
mode. The wave-vector, $\bf{k}$, of this mode can be assumed to be
parallel to the $z$ direction, so we adopt a metric where the $x$ and
$y$ directions are homogeneous and the inhomogeneous contributions
are functions of $z$ alone.  The simplest choice of metric which
expresses this constraint is
\be \label{metric}
ds^2 = dt^2 - A^2(t,z)\,dz^2 - B^2(t,z)\,(dx^2 + dy^2).
\ee
A single scalar field has the Lagrangian
\be \label{lagr}
{\cal L} = \ha \Pt^2 - \frac{\Pz^2}{2  A^{2}}- V(\phi)
\ee
and equation of motion
\be \label{eomphi}
\Ptt = 
   \frac{\Pzz}{A^2} -\left(\frac{\At}{A} + 2\frac{\Bt}{B}\right)\,\Pt
+ \left( \frac{2\Bz}{B A^2} -\frac{\Az}{A^3}
   \right)\,\Pz - \frac{d{V}}{d{\phi}}.
\ee
Overdots indicate derivatives with respect to $t$, and primes with respect
to $z$.  The non-zero components of the Einstein field equations,
$G_{\mu\nu} = -\kappa^2 T_{\mu\nu}$, are therefore
\begin{eqnarray} \label{eomA} 
\Att &=& -\frac{\At\Bt}{B} + \frac{A\Bt^2}{2B^2} + \frac{\Bzz}{AB} -
\frac{\Bz^2}{2AB^2} - \nonumber \\
& & \frac{\Az\Bz}{A^2B} -\frac{ \kappa^2 A}{2} \left[\frac{ \Pt^2}{2}
-\frac{3\Pz^2}{2 A^{2}}   - V \right], \\
  \label{eomB} 
\Btt &=&  \frac{\Bz^2}{2A^2B} -\frac{\Bt^2}{2B} -  
  \frac{ \kappa^2 B}{2} \left[\frac{\Pt^2}{2} + \frac{\Pz^2}{2 A^2}-V 
\right],\\
\label{constr1}
\frac{\Btz}{B}  & =& \frac{\At\Bz}{AB} -\frac{\kappa^2}{2} \Pt\,\Pz ,\\
\label{constr2}
\frac{\Bt^2}{2B^2} &=& \frac{\At\Bt}{AB} + \frac{\Bzz}{A^2B} +
\frac{\Bz^2}{2A^2B^2} - \frac{\Az\Bz}{A^3B} + \nonumber \\
& & \frac{ \kappa^2}{2} \left[\frac{\Pt^2}{2} + \frac{\Pz^2}{2 A^2} +
V \right]. 
\end{eqnarray}
\Eqs{constr1}{constr2} are constraints which must be satisfied by the
initial conditions, and are then used to check the accuracy of our
numerical integration. 

We choose $A$, $B$ and $\phi$ to be independent of $z$ at $t=0$, so
the initial inhomogeneity is expressed by $\At$, $\Bt$ and $\Pt$. This
makes it easier to ensure that the initial data satisfies the
constraints, but does not involve an unacceptable loss of generality.
With $\Az(0,z) = \Bz(0,z) = \Pz(0,z) = 0$ the first constraint,
\eq{constr1}, is trivial.  Rescaling the spatial co-ordinates fixes
$A(0,z) = B(0,z) =1$, while we denote $\phi(0,z) = \phi_0$.  
\Eq{constr2} requires
\begin{eqnarray}
\At(0,z) &=& \frac{\kappa^2}{2C}\left( \frac{\Pt^2}{2} + V(\phi_0) \right) 
- \frac{C}{2}, \\
\Bt(0,z) &=& 
\sqrt{\frac{\kappa^2}{3} \left( \frac{\langle\Pt^2\rangle}{2} +
V(\phi_0)\right)}= C , 
\end{eqnarray}
where we pick $C$ so that $\langle \At(0,z) \rangle = \langle \Bt(0,z)
\rangle$, where $\langle \cdots \rangle$ is a spatial average.  This
normalizes the metric functions and scales the co-ordinates, while
$\Pt(0,z)$ determines the configuration of the primordial universe
when we begin our simulation.  All simulations presented here begin at
the moment when inflation ends in the unperturbed model, defined by
$\ddot{a}(t) =0$, where $a(t)$ is the scale factor of the unperturbed
universe.

To facilitate comparison with the perturbative result, we excite a
single $k$-mode
\be \label{pert1}
\Pt(0,z) =  \Pt_0 + \epsilon \sin{\left(\frac{2\pi k z}{Z}\right)},
\ee
where $Z$ is the length of our ``box''. This definition rescales $k$
in units of $Z$, and we use this form throughout the rest of this paper.
The magnitude of the perturbation is fixed by $\epsilon$. Most
inflationary models predict that $\epsilon \ll 1$, but we can consider
an arbitrary $\epsilon$, since we are working with the full Einstein
equations.

We use solve the field equations on an $N$ point spatial grid. The
simulations presented here all have $N=1024$. We assume periodic
boundary conditions, and use a fourth-order differencing scheme to
express \eqs{eomA}{eomB} as a set of $N$ ordinary differential
equations, which are solved with a fourth-order Runge-Kutta
integrator. The resulting system suffers from a buildup of
short-wavelength numerical noise, common in {\it free-evolution\/}
schemes such as ours, due to the discretization process which can only
approximate the true time-like hypersurface from timestep to timestep
\cite{Choptuik1991a}. To control this we use a pseudo-spectral
approach, {\it de-aliasing\/} \cite{GottliebBK1}. This is implemented
by filtering $A$, $B$, $\phi$ (and their associated velocities),
removing Fourier components with wavelengths less than 8 grid spacings
at each timestep. Thus our effective spatial resolution is $\sim 16$
times larger than the separation of the individual points, $\Delta z$.

\vskip 5mm

\noindent {\bf III. Testing Perturbation Theory} Our first goal is to test
the perturbative analysis of the interaction between the field and
background perturbations in an inflationary model with the potential,
$V(\phi) = m^2 \phi^2 /2$
\cite{KodamaET1996a,NambuET1996a,BassettET1998a,FinelliET1998a}. In
particular, Finelli and Brandenberger \cite{FinelliET1998a} showed
that first order gravitational perturbation theory does not predict
the parametric amplification of scalar perturbations with wavelengths
considerably larger than the Hubble radius, as was suggested by
Bassett {\it et al.\/} \cite{BassettET1998a}.

To independently check Finelli and Brandenberger's prediction and to
confirm that perturbation theory gives an accurate description of the
universe at the end of inflation, we use our nonlinear code to analyze the
same system.  Consequently, we must relate our metric, which corresponds
to a particular gauge choice, to gauge invariant cosmological
perturbations. 

Working in physical time, with our symmetry constraint and choice of
flat spatial sections in the unperturbed universe, the general scalar
perturbation is \cite{MukhanovET1992b},
\begin{eqnarray}
ds^2 &=& (1+2\varphi(t,z))dt^2 - 2a(t)b(t,z)_{,i} dx^i dt \nonumber \\ 
& & - a^2(t)\left[(1-2\psi(t,z))\delta_{ij} + 2 E(t,z)_{,ij}\right] dx^i dx^j
\end{eqnarray}
where commas denote ordinary derivatives. The background solution,
$a(t)$, is defined by the FRW universe with our chosen initial
conditions and $\epsilon=0$.  Comparing with \eq{metric}, we deduce
that
\begin{eqnarray}
&\varphi = 0,\qquad b' = 0& \\
&A^2 = a^2(t) (1-2\psi + 2E'') &,\\
&B^2 =  a^2(t) (1-2\psi).&
\end{eqnarray}
We are now in a position to calculate the gauge invariant metric
perturbation $\Phi$,
\be 
\Phi = \varphi + \frac{\partial}{\partial t} \left[a (b- a\dot{E})\right]   
  = -\frac{\partial }{\partial t}(a^2 \dot{E}),
\ee
where we have set integration constants in the final version to ensure
that $\Phi$ vanishes when there is no perturbation.  We can express
$E''$ and, therefore, $\Phi''$ in terms of the metric functions and
their derivatives
\begin{eqnarray}
\Phi'' &=& \Btt B + \Bt^2 - \Att A - \At^2 + \nonumber \\
  & & \dot{H}(A^2 - B^2) + 2H(\At A -\Bt B),
\end{eqnarray}
where $H= \dot{a}/{a}$. Perturbation theory predicts the Fourier
modes, $\Phi_k$, of $\Phi$. Recalling the Fourier transform,
$\cal{F}$, of a derivative, we obtain $\Phi_k = -k^{-2}
{\cal{F}}(\Phi'')_k$. However, in linear perturbation theory $\Phi_k$
is obtained from Mukhanov's variable, $Q$ \cite{Mukhanov1988a}, which
satisfies \cite{MukhanovET1992b,KodamaET1996a,NambuET1996a}
\begin{eqnarray}
Q_k &=&\delta \phi_k + \frac{\dot{\phi}}{H} \Phi_k, \\
\ddot{Q_k} &+& 3H\dot{Q_k} + \nonumber \\
&& \left[\frac{k^2}{a^2} + \frac{d^2V}{d\phi^2}+
  2\frac{d}{dt}\left(\frac{\dot{H}}{H} + 3H\right)\right]Q_k=0, \\
\frac{k^2}{a^2} \Phi_k &=& \frac{\kappa^2}{2} \frac{\dot{\phi}^2}{H}
\frac{d}{dt}\left( \frac{H}{\dot{\phi}} Q_k \right) .
\end{eqnarray}

\begin{figure}[tbp]
\begin{center}
\begin{tabular}{c}
\epsfysize=4cm 
\psfrag{x}[bt]{$t$}
\psfrag{y}[tb]{$\mbox{Im}(\Phi_1)$}
\epsfbox{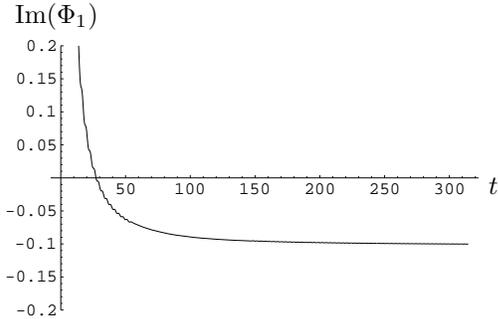}
\end{tabular}
\end{center}
\caption[fig1]{The evolution of the $\Phi_1$ mode, with a wavelength
$Z$, is plotted for $\epsilon=0.001$, After an initial transient
decays the solution tends towards a constant, as expected for a
super-horizon mode. The time is normalized so that $m=1$. For our
choice of phase, only the imaginary part of $\Phi_1$ is excited.
\label{perturb1}}
\end{figure}

We ran our code for an initial perturbation with a wavelength equal to
the size of our simulation region, which is $10^2$ larger than the
post-inflationary Hubble length, and $\epsilon=.001$. This value of
$\epsilon$ is somewhat arbitrary, but corresponds to a density
contrast of roughly the same magnitude as those observed at COBE
scales.  We extracted $\Phi$, as described above, and solved
numerically for $Q$ with initial conditions chosen to match those of
the nonlinear simulation.  We find that the nonlinear and the
perturbative results agree to within approximately 1 part in
$10^5$. The evolution of $\Phi_k$ corresponding to the single excited
mode is shown in Fig~1.  This verifies both the perturbative treatment
of the post-inflationary era in the absence of resonance {\it and\/}
the specific conclusions of Finelli and Brandenberger
\cite{FinelliET1998a}, {\it i.e.\/} that for $m^2\phi^2$ inflation,
$\Phi_k$ is constant for super-horizon modes and does not undergo
parametric amplification.

\vskip 5mm

\noindent {\bf IV. Beyond Linear Gravity}
We now consider the role of nonlinear gravitational effects during
preheating. A full relativistic treatment (or higher order
perturbation theory) couples Fourier modes which are independent at
first order.  Post-inflationary perturbations are small, but in
$\phi^4$ inflation (and in more complicated models), parametric
resonance can increase $\Phi$ dramatically. If this growth lasts long
enough, nonlinear gravitational effects are unavoidable.

We leave the full numerical analysis of models with more complicated
potentials for future work, and use our $m^2 \phi^2$ system as a toy
model to investigate the consequences of one or more modes acquiring a
large amplitude. Since we have just shown that super-horizon modes are
not amplified with a $\phi^2$ potential, we must inject this through
our initial conditions.

\begin{figure}[tbp]
\begin{center}
\begin{tabular}{c}
\epsfysize=4cm 
\psfrag{x}[bt]{$t$}
\psfrag{y}[tb]{$|\Phi_k|$}
\epsfbox{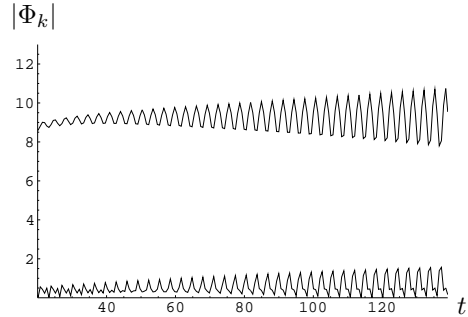}
\end{tabular}
\end{center}
\caption[fig2]{The evolution of $|\Phi_5|$ (upper) and $|\Phi_{10}|$ are
plotted, for the nonlinear case of $\epsilon =.1$. In this case, there
is a significant transfer of power to higher modes.  
\label{coupled1}}
\end{figure}

We examine a perturbation with $k=5$, and $\epsilon=0.1$, which is
large in contrast to the primordial COBE scale perturbations, but is
not inconsistent with a mode amplified by parametric resonance.  Two
$\Phi_k$ are plotted in Fig.~2.  Modes for which $k$ is an integer
multiple of the $k$ of the initial perturbation are excited. The $k=5$
mode is outside the initial Hubble radius, but nonlinear effects cause
it to grow, and the perturbative prediction is not applicable. We also
observe that in initially over-dense (with respect to a flat,
unperturbed universe) regions, $A$ begins to decrease, signifying the
onset of gravitational collapse.

\begin{figure}[tbp]
\begin{center}
\begin{tabular}{c}
\epsfxsize=8.55cm 
\psfrag{k1}[]{$|\Phi_1|$}
\psfrag{k9}[]{$10^3 |\Phi_9|$}
\psfrag{k7}[]{$100 |\Phi_7|$}
\psfrag{k8}[]{$|\Phi_8|$}
\epsfbox{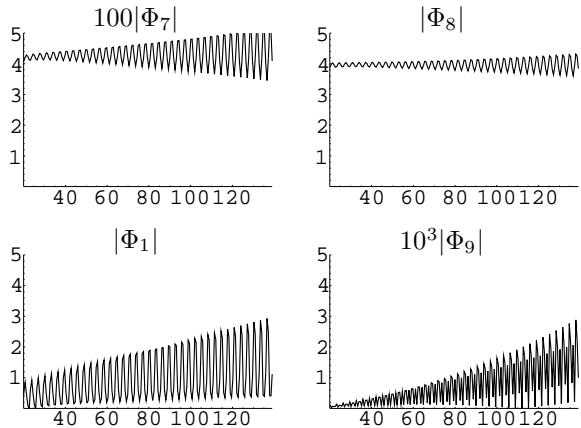}
\end{tabular}
\end{center}
\caption[fig3]{Several modes of $\Phi$ are plotted with
$\epsilon=0.05$ and $k=8$, and the perturbation given by
\eq{pert2}. Here, $\Phi_1$ is initially excited, but significant
amounts of power are transferred to other modes.
\label{coupled2}}
\end{figure}

Finally, we consider a more complex case where two modes, whose $k$
values are not multiples of one another, are initially excited.
Specifically, we replace \eq{pert1} by
\begin{eqnarray} 
\Pt(0,z) &=&  \Pt_0 + \epsilon \left[ \sin{\left(\frac{2\pi k
z}{Z}\right)} \right. \nonumber \\ & & \left. + 0.01
\sin{\left(\frac{2\pi (k-1) z}{Z} + \frac{\pi}{4}\right)}\right].
\label{pert2}
\end{eqnarray}
The two modes we initially excite become mixed in the gauge invariant
perturbation, so $\Phi_1$ always has a significant amplitude. This can
be seen in Fig.~3, which corresponds to the choice $\epsilon = 0.05$
and $k=8$. However, unlike the previous case where only modes which
were multiples of the initially excited mode were enhanced, we now see
a transfer of power between the initially excited ``band'' and a wide
variety of other modes. Consequently, if some modes acquire a
significant amplitude through parametric resonance, nonlinear effects
can then transfer power to modes which fall outside the resonance
bands.

\vskip 5mm

\noindent {\bf V. Discussion} Nonlinear effects may be important in a
number of ways not considered here. For instance, when inflation is driven
by two interacting scalar fields, the homogeneous post-inflationary
universe may exhibit chaos \cite{EastherET1997a}, and the interaction
between this chaos and nonlinear inhomogeneous effects is intriguing. 
Likewise, parametrically enhanced perturbations will grow further due to
nonlinear gravitational effects and could eventually undergo gravitational
collapse, leading to the formation of primordial black holes in models
where their existence is not currently predicted\cite{BassettET1998a}. The
scattering of scalar perturbations is also a nonlinear effect and will
lead to gravitational wave
production\cite{KhlebnikovET1997c,Bassett1997,BassettET1998a}.

By abandoning perturbation theory and solving the Einstein field
equations directly, we have laid the foundation for a fully nonlinear
analysis of preheating in an expanding universe, which will include
all possible couplings between perturbations in spacetime and the
classical field(s). We verified Finelli and Brandenberger's analysis
of $m^2\phi^2$ inflation, showing that the super-horizon modes of the
gauge invariant perturbation are constant, and unamplified by any
resonant processes. We also made a preliminary exploration of
interactions between modes when the perturbation is large,
demonstrating the ability of nonlinear effects to couple modes, which
allows a transfer of power from modes which are parametrically
enhanced to those which are not in resonance at the perturbative
level. Consequently, we conjecture that the nonlinear gravitational
effects will broaden any resonance bands seen in the conventional
perturbative analysis; we are currently investigating these questions
in detail.

\vskip 5mm

{}\noindent {\bf Acknowledgments} We thank Robert Brandenberger, Fabio
Finelli, David Kaiser, Mark Trodden and Tim Warburton for useful
discussions.  Computational work in support of this research was
performed at the Theoretical Physics Computing Facility at Brown
University.  RE is supported by DOE contract DE-FG0291ER40688, Task
A.

\end{document}